\documentstyle[aasms]{article}
\tightenlines
\begin{document}

\title{Thick to Thin: The Evolutionary Connection Between PG 1159 Stars and
the Thin Helium-Enveloped Pulsating White Dwarf GD 358}

\author{Benjamin T. Dehner\altaffilmark{1} and
Steven D.  Kawaler\altaffilmark{1}}

\altaffiltext{1}{Dept. of Physics and Astronomy,
Iowa State University, Ames IA, 50011; email: btd@iastate.edu}

\begin{abstract}

Seismological observations with the Whole Earth Telescope (WET) allow the
determination of the subsurface compositional structure of white dwarf
stars.  The hot DO PG~1159--035 has a helium surface layer with a mass of
$\sim 10^{-3} M_{\odot}$, while the cooler DB white dwarf GD~358 has a much
thinner surface helium layer of 10$^{-6} M_{\odot}$.  Taken literally, these
results imply that either there is no evolutionary relation between these two
stars, or that there is an unknown mass loss mechanism.  In order to
investigate a possible evolutionary link between these objects, we computed
evolutionary sequences of white dwarf models that included time-dependent
diffusive processes.  We used an initial model based on the PG~1159
pulsational data, which has a surface layer $3\times 10^{-3}M_{\odot}$ thick,
and a composition of 30\% helium, 35\% carbon, and 35\% oxygen.  Below this
surface layer is a thin transition zone where the helium fraction falls to
zero.  As expected, diffusion caused a separation of the elements; a
thickening surface layer of nearly pure helium overlays a deepening transition
zone where the composition changes to the surface composition of the original
model.  When the model reached the temperature range inhabited by GD~358 and
the pulsating DB white dwarfs, this pure helium surface layer was $\sim
10^{-5.5}M_*$ deep.  The resulting evolved model is very similar to the model
used by Bradley and Winget (1994) to match the pulsation observations of GD
358.  The pulsation periods of this model also show a good fit to the WET
observations.  These results demonstrate the plausibility of a direct
evolutionary path from PG~1159 stars to the much cooler DB white dwarfs by
inclusion of time-dependent diffusion.  A problem still remains in that our
models have no hydrogen, and thus must retain their DB nature while evolving
through the T$_{\mbox{eff}}$ 45,000K to 30,000K.  Since there are no known DB
stars in this range, we plan to address this problem in future calculations.

\end{abstract}

\subjectheadings{diffusion --- stars: evolution --- stars: individual (GD
358, PG 1159) --- white dwarfs}

\section{Introduction: The Problem of Ancestry}

Seismological exploration of white dwarf stars with the Whole Earth
Telescope, described by Nather et al. (1990), has yielded unprecedented
details about their subsurface compositional stratification.  Winget et
al. (1991) observed the pulsations of the hot ($T_{\rm eff}$=140,000K) white
dwarf PG 1159--035 (hereafter simply ``PG~1159''),
uncovering over 120 independent pulsation modes.  These modes
appear at frequencies that are naturally explained by pulsation theory
as nonradial $g-$modes.  Kawaler \& Bradley (1994: ``KB'') examined the
observed pulsation periods in detail, and determined several parameters of
this star such as its mass and luminosity.  In addition, they show that
PG 1159 has a subsurface composition transition zone at about
$3\times 10^{-3} M_{\odot}$ below the surface.  The surface composition
of that star is roughly 33\% He, 50\% $^{12}C$, and 17\%$^{16}O$ by mass,
(Werner et al. 1991); the composition transition is where the helium mass
fraction goes down to zero.  Thus the surface layer contains roughly
$10^{-3} M_{\odot}$ of helium.  A likely route of evolution to the PG~1159
phase begins with departure from the AGB during a thermal pulse (Iben \&
Tutukov 1984).  Such a model has at most $10^{-2}M_{\odot}$ of helium in the
surface layers.  With modest mass loss between the AGB and the PG~1159 stage,
the helium layer mass determined by KB fits this picture.

Winget et al. (1994) report observations of the pulsating white dwarf,
GD~358.  This star is much cooler than PG~1159, and is the prototype of the DB
pulsators, with an effective temperature of about 25,000K (Thejl, Vennes, and
Shipman 1991).  Winget et al. (1994) found over 180 separate pulsation
frequencies in this star.  Model analysis of this star by Bradley \& Winget
(1994: ``BW'') successfully reproduced in detail the pulsation frequencies
with a model of about the same mass as PG~1159, but with a surface layer
of pure helium of approximately $1.2\times 10^{-6} M_{\odot}$.  This is three
order of magnitude smaller than PG~1159.

This difference in helium layer mass suggests that establishing an
evolutionary connection between these two objects requires finely--tuned mass
loss beyond the PG~1159 stage.  Alternatively, these results challenge the
notion of there being a direct evolutionary relationship between these
objects.

The models used by KB were evolutionary models which had a single common
ancestor on the AGB.  The time that they took to reach the PG 1159
stage was less than $10^6$ yr.  The models used by BW were also evolutionary
models with ages of several $\times 10^7$ years after departure from the AGB.
In both calculations, the compositional stratification was fixed at
the start, and did not change in the course of the calculation.  However,
diffusion by gravitational settling, acting on reasonably short time scales,
is believed to be responsible for the extremely pure surface compositions
of most white dwarfs stars (Schatzman 1958).  While there has been much
discussion of diffusion in the literature, there have
few attempts to include time--dependent diffusion as part of an evolutionary
calculations (see for example Iben and MacDonald 1985).  Other white dwarf
evolution calculations that address diffusion usually assume that diffusion
reaches equilibrium quickly, and so evolve models with equilibrium diffusion
profiles in their interiors (e.g. Tassoul, Fontaine, \& Winget 1990).  The
time scale to reach equilibrium increases dramatically with depth and with a
shallow gradient.  In realistic models, diffusion takes a long time to
approach equilibrium in deeper layers, and equilibrium concentrations are
reached from the outside inward.  Therefore, the use of equilibrium (or
near-equilibrium) diffusive profiles is open to question in objects as young
as PG~1159 and GD~358.  BW's evolutionary models employ an equilibrium
diffusion profile above the midpoint of the composition transition region,
and enforce a steeper gradient below, which crudely models the approach to
diffusive equilibrium in deeper layers.

This paper reports on results of our calculations of time--dependent diffusion
in evolving white dwarf models.  Since diffusive equilibrium is
only truly achieved after an infinite time, we constructed a series of
evolutionary models including time--dependent diffusive processes.
We find that GD~358 is a ``snapshot'' of diffusion in progress; it
represents an intermediate step in the approach to diffusive
equilibrium.  Its outer layer structure is a natural consequence of the
evolution of the compositional structure of a PG~1159-like star.  In
Section II below we briefly describe the implementation of diffusion within
our white dwarf evolution code.  In Section III, we show the results of
evolution of models with starting models taken from KB , and compare the
results to the models used by BW.  Since pulsation periods and their
differences are the prime observational test of the model, in Section IV we
compare the pulsation properties of our models with GD 358.  Section V
concludes this paper with a discussion of the evolutionary link between the
PG 1159--035 stars and the DB stars.

\section{Evolutionary Models of White Dwarfs with Time--Dependent Diffusion}

The computer code we used was a standard stellar evolution code
(ISUEVO; Dehner, 1995) which was optimized for use in pulsation studies.
This code was modified to include time--dependent diffusion in the
following way. The equation describing diffusion is
\begin{equation}
\frac{\partial n_i}{\partial t} = -\frac{1}{r^2}\frac{\partial (r^2 w_i n_i)}
{\partial r},
\end{equation}
where $n_i$ is the number density of species $i$ (element or  electrons), $t$
is time, $w$ is the diffusive velocity, and $r$ the radial coordinate.
The code solves equation [1] using a technique known as the method of lines.
First, equation [1] is spatially discretized using centered differences.  Each
evolutionary time step is then broken down into substeps.  Within each
substep, the set of discretized diffusion equations is integrated for each
element using the backward differentiation integrator DEBDF by Shampine and
Watts (1979), after which the abundances are updated.  Typically, the first
few substeps are chosen to be very short (5 years), with subsequent steps
being determined by considering the maximum composition change at all zones
from the previous step.

The diffusive velocity $w_i$ is obtained using
the following equation derived from Burgers (1969)
\begin{equation}
\label{eq:diffv}
\nabla p_i - \frac{\rho _i}{\rho} \nabla p - n_i Z_i e {\bf E} = \sum _j^{M+1}
K_{ij}({\bf w}_j - {\bf w}_i),
\end{equation}
where $i$ and $j$ are species indices, $Z$ is the electric charge, ${\bf E}$
is the electric field, and $K_{ij}$ is the resistance coefficient, and all
other symbols have their usual meanings.  The resistance coefficients,
inversely related to the diffusion coefficients used by some authors, are
obtained from the tables of Paquette et al. (1986).  Equation [2]
follows from Burgers (1969) after assuming that thermal diffusion and
magnetic field effects are negligible (cf. Burgers 1969, eqn. [18.1].)

We also ignore the effect of radiative forces.  The principal aim of this
paper is to explore time--dependent diffusion of dominant species.
Radiative forces might result in modest increases in the abundances of
heavy elements, but probably would not significantly modify the helium
abundance in terms of the pulsation diagnostics.  We recognize that
radiative levitation studies are very fashionable in the white dwarf
literature because of the importance of heavy element effects on white
dwarf fluxes.  Clearly, radiative forces are important for hot white dwarfs
such as those detected by EUVE, ROSAT, et al.; we are currently working
on including simple radiative levitation to address other issues in white
dwarf evolution.

We evolved four evolutionary sequences with masses from 0.56$M_{\odot}$ to
0.62$M_{\odot}$, to bracket the mass determined by BW for GD 358.  The
starting model for these sequences was the $0.59M_{\odot}$ static model which
best reproduced the pulsation spectrum of PG~1159 from KB.  This model, with
$T_{\rm eff}=140,000K$, had a surface layers with uniform composition of 30\%
helium, 35\% carbon, and 35\% oxygen.  In a transition region at $3\times
10^{-3}M_*$ below the surface, the composition changed to C=O=50\%.  The
thermal structure of this model is that of an evolutionary model of a
post-AGB star which left the AGB during a helium shell flash.  To obtain
starting models at different masses, we simply re-integrated the model with
the chosen mass so as to reach a consistent central boundary condition.  This
procedure results in realistic models that closely match evolutionary models
with similar masses and histories.  For a comparison between these starting
models and full evolutionary models, see KB.  Evolution proceeded into the
temperature domain of the DB variables; when convection developed in these
models we computed the convective flux using the ML2 formulation of Tassoul,
Fontaine, \& Winget (1990).

\section{Model Results: The Evolutionary Connection}

The results of the diffusion calculations reveal, as expected, that carbon
and oxygen sink downward, and helium floats to the top.  Figure 1 shows the
helium mass fraction as a function of the stellar mass fraction below the
surface ($q_s$) for several models in the 0.58$M_{\odot}$ sequence.  Models
in the other sequences evolve in very similar ways.  The starting model shows
a constant helium abundance down to $\log q=-3.1$; the helium abundance drops
to zero at $\log q=-2.6$.  After 400,000 years, the model reaches an
effective temperature of 76,000$K$. At this point, a steep composition
gradient has developed at $\log q=-8.0$.  Below this gradient, the abundance
drops gradually down to the original surface helium abundance of 30\%.  The
initial composition transition at $\log q=-2.6$ is essentially unchanged.  At
$2\times 10^6$~yr after the start ($T_{\rm eff}=46,100K$), the outer
composition transition zone lies at $\log q=-7.5$.  Once the model reaches
the $T_{\rm eff}$ range where pulsating DB white dwarfs are found (in Figure
1, this is the model at $T_{\rm eff}=28,600K$) at an age of $8.8\times 1
0^6$yr,  the outer composition transition has reached down to $\log q=-6.5$.
As the model evolves through the instability strip, this He/C transition zone
deepens further, reaching $\log q=-6.0$ at $T_{\rm eff}\sim 24,000K$ ($2.0
\times 10^7$yr), and deepening to $\log q=-5.6$ at $T_{\rm eff}\sim 21,000K$
($4.1 \times 10^7$yr).  Even in these later models, diffusion has only
slightly broadened the zone where the helium abundance drops from the
original surface value to zero.  This is to be expected because of the very
long diffusion time scales that deep in the star.

The models of BW were constructed using an entirely different evolutionary
code and input physics, we compare the global parameters of our models with
their best model for GD 358 to see how similar the models are.  Bradley
(private communication) reports that their best model had $M=0.61M_{\odot}$,
$T_{\rm eff}=24,210$, $\log(L/L_{\odot}=-1.3077$, and $R=0.01275R_{\odot}$.
The closest model in our $0.60M_{\odot}$ sequence was at $T_{\rm
eff}=24,172$; that model had a luminosity of $\log (L/L_{\odot})=-1.3125$ and
a radius of 0.01263$R_{\odot}$.  Most significantly, the composition
transition zone lies at the same position below the surface in both models.
Thus, the global parameters for the diffusion models and the models of BW are
quite similar.  The evolutionary sequence that we calculated produced a model
that is extremely close to the model of BW that was designed to fit the
observed properties of GD~358.

In our models, the starting point was a tested stellar model that reproduced
the observations of a much younger object than GD~358.  Therefore, evolution
of these stars leads to a natural and direct evolutionary connection between
the prototype of the PG~1159 stars and the prototype of the DB pulsators.

\section{Testing the Connection}

BW used the pulsation periods observed in GD 358 to constrain the
compositional structure of the outer layers of the star.  The observation
that demanded a thin helium layer was the nonuniformity of the spacing
between successive overtone periods in that star.  Through the effects of mode
trapping (KB, Brassard et al. 1992, Bradley, Winget \& Wood 1992), these
departures occurred regularly, with the number of modes between minima in the
spacing increasing with decreasing surface layer thickness.  For GD~358, the
fit was to a single pronounced minimum in the period spacing, because of the
limited number of certain overtone periods seen by Winget et al. (1994).

To see how well our models match the observed pulsation properties of
GD~358, we computed pulsation periods for all models in all four sequences
that were in or near the DB instability strip.  We then compared the computed
periods to those seen in GD~358 by Winget et al. (1994), and looked for those
models which most closely matched the star.  We used two quantitative
comparisons: the periods themselves, and the differences between successive
periods.  The mean differences between the observed and computed periods (and
the period spacings) reaches a minimum for the models that are the closest
fits to GD~358.  It should be stressed that the periods and the period
spacings are two separate criteria for comparing models with observations.

To compare each model with observations, we computed the sum of the squares
of the differences between the periods of each model and GD~358, and did
the same for the period spacings.  Figure 2 shows the inverse
sum--of--squares difference in the periods, and in the period spacings, as a
function of effective temperature for the three sequences bracketing the
BW value of $0.58M_{\odot}$.  Remarkably
the minimum difference (which appear as peaks in this plot) for
both the periods and the period spacings are very close together for these
sequences.  The overall best model is the $0.58M_{\odot}$ model at $T_{\rm
eff}=24,121$.  The mean period spacing of this model is $39.52 \pm 3.2 s$,
compared to the observed value of $39.5 \pm 5.2$.  In comparison with the
models of BW, while they have an overall better fit to the observed pulsational
properties of GD~358, our models have the advantage of not being tuned
to fit the observations, but were evolved directly from the PG~1159 stage.

\section{Conclusions: An Evolutionary Link Between DBV Stars and PG 1159
Stars}

Our models demonstrate how chemical diffusion can cause PG~1159 stars to
evolve into DBV stars such as GD~358 without appeal to mass loss, despite the
incongruity in surface helium layer masses.  This gives one definitive path
of evolution of white dwarf stars from the PG~1159 stage, and is a step in
clarifying white dwarf evolutionary relationships.  This also demonstrates
the necessity of invoking time-dependent diffusion when considering white
dwarf evolution.

In this investigation, we assumed a zero hydrogen fraction in all of our
models.  This implies that that they must retain their DB nature while
cooling through the DB gap from 45,000 K to 30,000 K.  This is an obvious
contradiction with the observed lack of DB stars in this temperature range
(Wesemael et al. 1985; Liebert 1986).  In future calculations with trace
hydrogen present, we will examine the behavior of the hydrogen and helium
layers in this range.

\acknowledgements

This work was supported by NSF Grant AST-9115213 and NSF Young Investigator
Award AST-9257049 to Iowa State University.  We also thank Ed Nather for his
continual insistence on quantifying the comparison of models with the
observations.


\newpage
\centerline{FIGURE CAPTIONS}

\begin{description}

\item[Fig.~1] Composition profiles for various models.  The x axis is the
log of the surface mass fraction, given by $q_s = (M_* - M_r)/M_{\odot}$.

\item[Fig.~2] Inverse sum--of--squares comparison of pulsation periods and
period spacings to GD 358 observations for all models in the DB instability
strip.

\end{description}
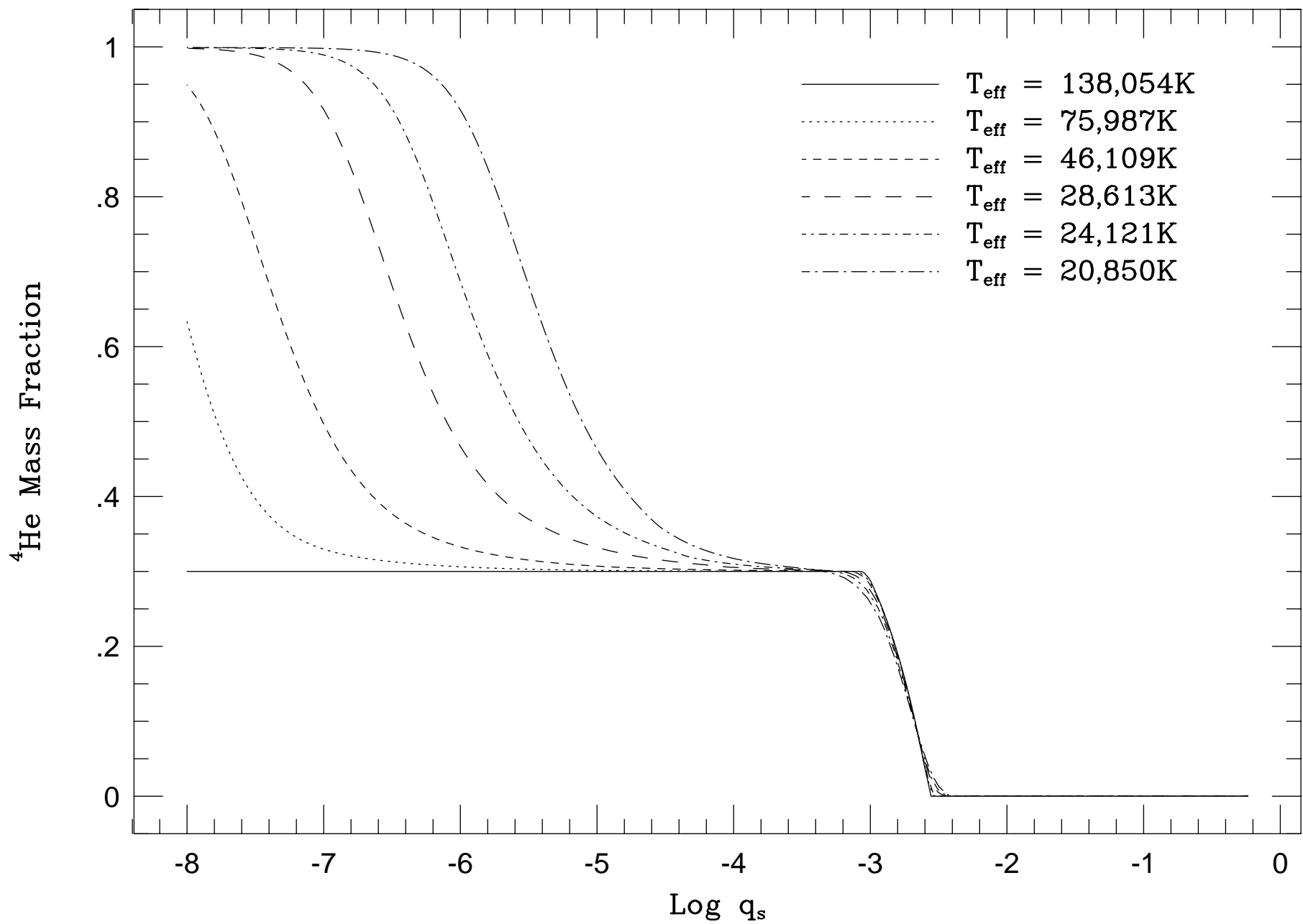
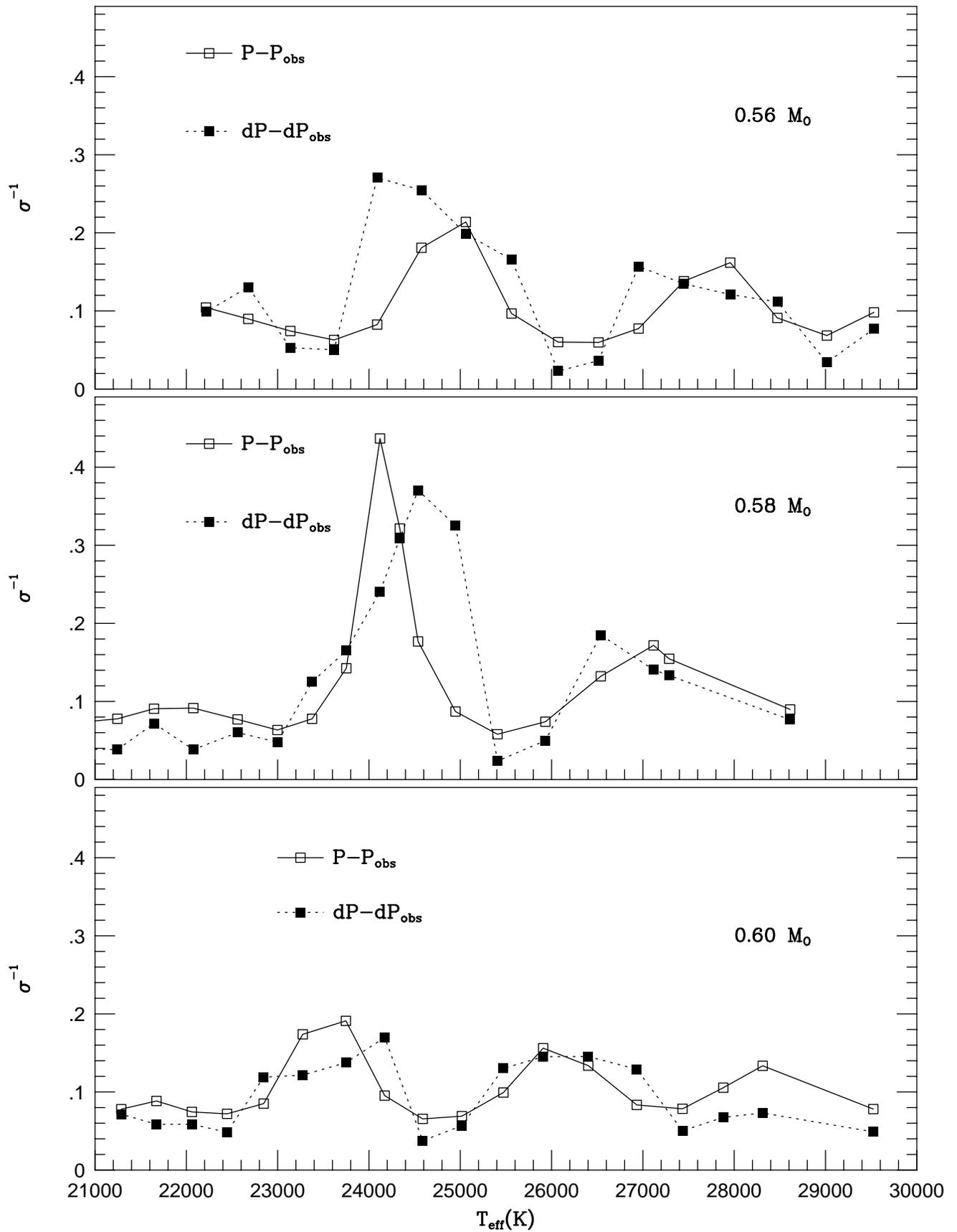
\end{document}